\documentclass[a4paper,11pt]{article}
\usepackage{pos}
\usepackage{graphicx}
\newcommand{\dd}{\mathrm{d}}

\usepackage{xcolor}

\title{Quantum Fourier integration of decay rates at next-to-leading order}
\author*[a]{David F. Rentería-Estrada}
 
\affiliation[a]{Instituto de F\'isica Corpuscular, \\ Universitat de Val\`encia--Consejo Superior de Investigaciones Científicas,\\ E-46980 Paterna, Spain}

\emailAdd{david.renteria@ific.uv.es}
\abstract{We discuss the use of quantum neural networks (QNNs) as a  tool for integrating decay rates at next-to-leading order (NLO) in perturbation theory. Data-encoding quantum circuits admit a truncated Fourier-series representation, enabling a trained QNN to approximate a target integrand and provide the coefficients subsequently integrated with Quantum Fourier Iterative Amplitude Estimation (QFIAE). The causal-unitary formulation of the Loop--Tree Duality (LTD) combines virtual and real contributions into locally finite integrands suitable for this representation. We consider $H\to q\bar q(g)$, $\gamma^*\to q\bar q(g)$ and $\Phi\to\phi\phi(\phi)$, comparing quantum-simulator and quantum-hardware results with analytic NLO predictions. The study illustrates the potential of Fourier-based QNN integration for high-energy physics.}

\FullConference{43rd International Conference on High Energy Physics (ICHEP 2026)\\
30 July  to 5 August , 2026\\
Natal, Brazil\\}

\begin{document}
\maketitle

\section{Introduction}
Quantum neural networks (QNNs) provide a useful representation of multidimensional functions for numerical integration. In particular, a variational quantum circuit with suitable data-encoding gates can be expressed as a truncated Fourier series: its accessible frequencies are determined by the encoding, while its coefficients depend on trainable circuit parameters~\cite{schuld,reupload}. This structure motivates the use of QNNs as the central approximation tool in Quantum Fourier Iterative Amplitude Estimation (QFIAE)~\cite{qfiae}, where the integrand is learned by a QNN and the resulting Fourier components are integrated using iterative quantum amplitude estimation (IQAE). Other methods include Quantum Adaptive Importance Sampling (QAIS)~\cite{Pyretzidis:2025stx}.

We apply this methodology to total decay rates at next-to-leading order (NLO), following the calculation in Ref.~\cite{original}. The physical integrands are provided by causal unitary in the Loop--Tree Duality (LTD), where virtual and real contributions are assembled from residues of vacuum amplitudes and their infrared singularities cancel locally~\cite{vacuum,causal}. This combination yields low-dimensional, comparatively smooth functions suitable for Fourier approximation. Section~2 outlines the NLO decay-rate representation and its reduction of integration variables. Section~3 discusses the QNN Fourier representation and the IQAE integration step. Section~4 presents results for three decay processes, followed by conclusions in Section~5.
%%%%%%%%%%%%%%%%%%%%%%%%%%%%%%%%%%%%%%%%%%%%%%%%%%%%%%%%%%%%%%%%%
%%%%%%%%%%%%%%%%%%%%%%%%%%%%%%%%%%%%%%%%%%%%%%%%%%%%%%%%%%%%%%%%%

\section{Locally finite NLO decay-rate integrands}
\label{sec:ltd}
\begin{figure}[!t]
\centering
\includegraphics[width=0.52\linewidth]
{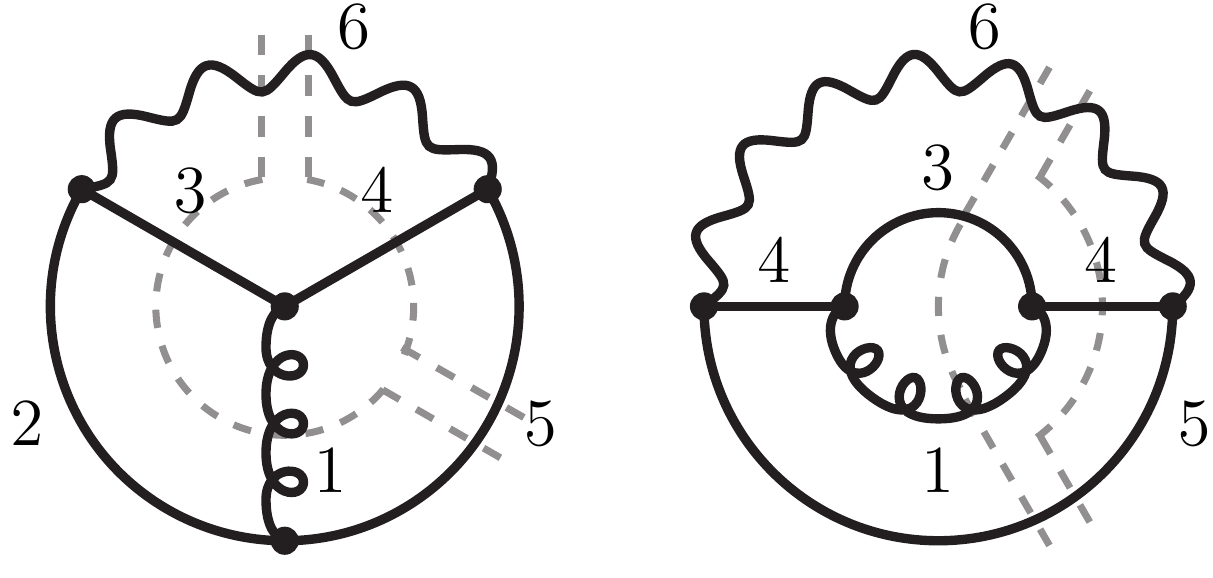}
\caption{
Three-loop vacuum diagrams contributing
to $\gamma^*\to q\bar q(g)$ at NLO.
The gray dashed lines identify phase-space
residues associated with different final
states. Analogous diagrams describe
$H\to q\bar q(g)$ and
$\Phi\to\phi\phi(\phi)$~\cite{original}.
}
\label{fig:diagrams}
\end{figure}

We consider the inclusive decays $H\to q\bar q(g)$, $\gamma^*\to q\bar q(g)$ and $\Phi\to\phi\phi(\phi)$, where the particle in parentheses denotes the additional real emission at next-to-leading order (NLO). The decay rate receives contributions from the interference between one-loop and tree-level amplitudes, as well as from tree-level processes with three particles in the final state. Within the causal-unitary LTD, these contributions are generated from phase-space residues of a common three-loop vacuum amplitude, illustrated in Fig.~\ref{fig:diagrams}~\cite{causal,original}. Following Ref.~\cite{original}, the differential NLO decay rate can be written as
\begin{equation}
\dd\Gamma_a^{(1)}
={}
\frac{\dd\Phi_{\boldsymbol\ell_1\boldsymbol\ell_2}}
{2\sqrt{s}}
\Big[
\mathcal A_{\rm D}^{(3,a,R)}(456)
\,\widetilde\Delta_{45\bar6} +
\mathcal A_{\rm D}^{(3,a)}(1356)
\,\widetilde\Delta_{135\bar6}
+
(5\leftrightarrow2,\,4\leftrightarrow3)
\Big],
\label{eq:ltd}
\end{equation}
where $\sqrt{s}$ is the mass of the decaying
particle. The residues
$\mathcal A_{\rm D}^{(3,a,R)}(456)$ and
$\mathcal A_{\rm D}^{(3,a)}(1356)$
represent the renormalised two-particle
virtual contribution and the three-particle
real-emission contribution, respectively.
The exchange of indices accounts for the
corresponding symmetric configurations.
The functions
$\widetilde\Delta_{i_n \cdots i_2 \bar i_1}=2\pi\delta(\lambda_{i_n \cdots i_2 \bar i_1})$
impose energy conservation for each
final-state configuration~\cite{causal,original}.
The integration measure is expressed in
terms of two independent loop
three-momenta,
\begin{equation}
\dd\Phi_{\boldsymbol\ell_1\boldsymbol\ell_2}
=
\prod_{j=1}^{2}
\frac{\dd^3\boldsymbol\ell_j}{(2\pi)^3}.
\label{eq:measure}
\end{equation}
In the rest frame of the decaying particle,
the integration measure depends on the
moduli of the two loop three-momenta,
$|\boldsymbol{\ell}_1|$ and
$|\boldsymbol{\ell}_2|$, and their relative
angle $\theta$. Rotational invariance
allows the integration over the remaining
angular variables. The relative angle is
parametrised as $\cos\theta=1-2v$,
with $v\in[0,1]$. The energy-conservation delta functions
fix one of the two radial variables:
$|\boldsymbol{\ell}_2|$ for the two-particle
virtual contribution, or
$|\boldsymbol{\ell}_1|$ for the three-particle
real-emission contribution. Consequently,
each contribution reduces to a
two-dimensional integral over $v$ and the
remaining independent radial variable,
$|\boldsymbol{\ell}_i|$, which is mapped
from $[0,\infty)$ to $[0,1)$ for numerical
integration. The corresponding phase-space
reduction and Jacobians are detailed in
Eqs.~(6)--(10) of Ref.~\cite{original}. Since the virtual and real-emission
contributions are expressed in common loop
variables, their infrared and threshold
singularities cancel locally when the
relevant terms are combined. Ultraviolet
divergences are removed through local
counterterms, yielding a finite integrand
that can be evaluated directly in the four physical space-time dimensions~\cite{causal,vacuum,original}.
This representation provides the target
function for the quantum neural network
and its subsequent truncated Fourier
decomposition.

%%%%%%%%%%%%%%%%%%%%%%%%%%%%%%%%%%%%%%%%%%%%%%%%%%%%%%%%%%%%%%%%%
%%%%%%%%%%%%%%%%%%%%%%%%%%%%%%%%%%%%%%%%%%%%%%%%%%%%%%%%%%%%%%%%%
%%%%%%%%%%%%%%%%%%%%%%%%%%%%%%%%%%%%%%%%%%%%%%%%%%%%%%%%%%%%%%%%%
\section{Quantum Fourier integration of NLO decay rates}
\label{sec:quantum}
\begin{figure}[t]
\centering
\includegraphics[width=0.75\linewidth]{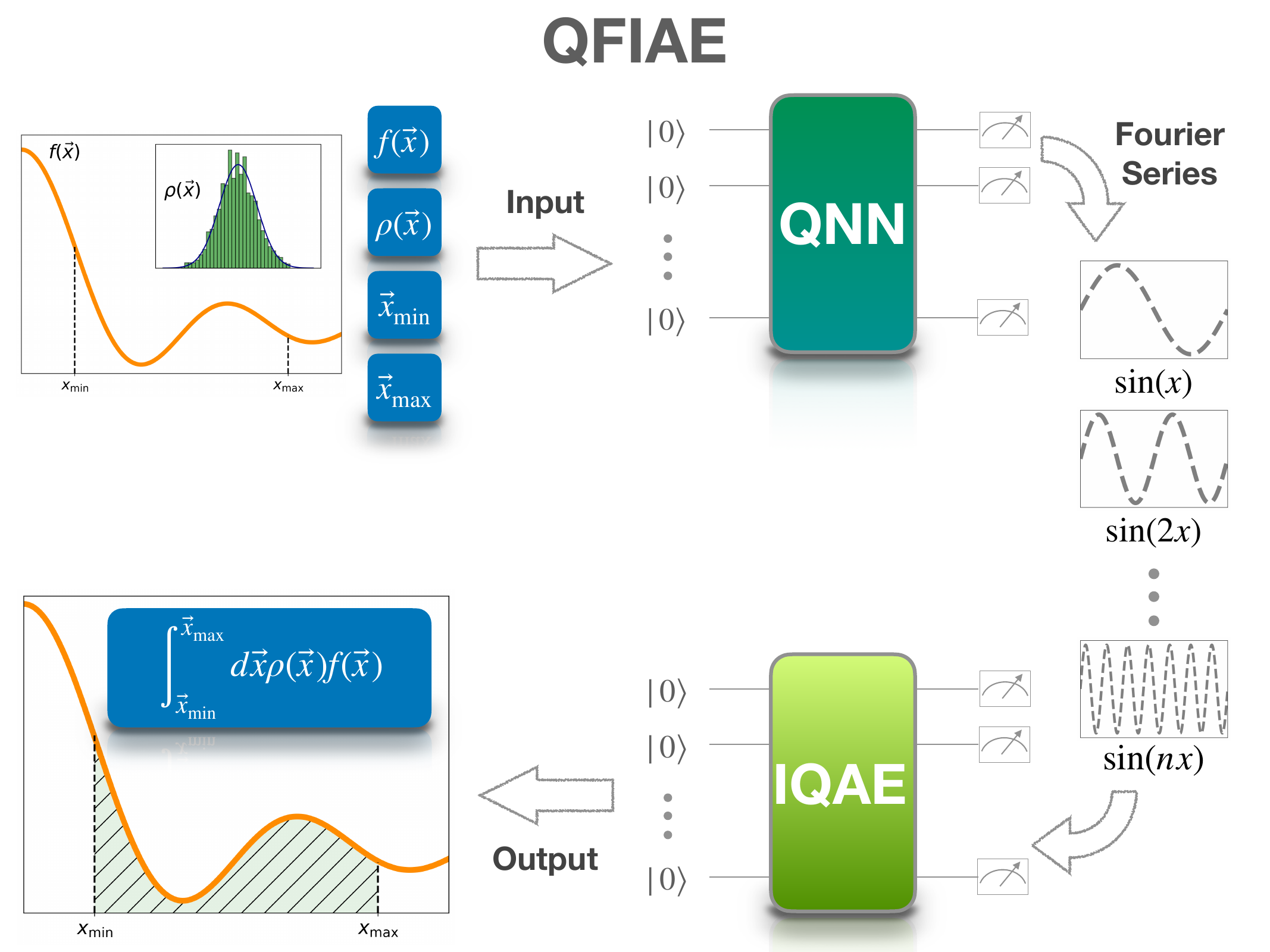}
\caption{
Schematic representation of the Quantum Fourier Iterative
Amplitude Estimation (QFIAE) algorithm. A quantum neural
network (QNN) approximates the target function through a
truncated Fourier series, whose coefficients are extracted
from the trained circuit. Iterative Quantum Amplitude
Estimation (IQAE) then evaluates the individual trigonometric
contributions, which are combined to reconstruct the
integral~\cite{original}.
}
\label{fig:qfiae}
\end{figure}

%%%%%%%%%%%%%%%%%%%%%%%%%%%%%%%%%%%%%%%%%%%%%%%%%%%%%%%%%%%%%%%%%
The locally finite NLO decay-rate integrands obtained through LTD causal unitary provide the input for the Quantum Fourier Iterative Amplitude Estimation (QFIAE) algorithm~\cite{original}. As illustrated in Fig.~\ref{fig:qfiae}, the method combines two main components: a quantum neural network~(QNN), which approximates the target integrand through a truncated Fourier representation, and Iterative Quantum Amplitude Estimation (IQAE), which evaluates the corresponding trigonometric integrals. The individual contributions are subsequently combined to reconstruct the total NLO decay rate.

%%%%%%%%%%%%%%%%%%%%%%%%%%%%%%%%%%%%%%%%%%%%%%%%%%%%%%%%%%%%%%%%%
\subsection{Fourier representation with quantum neural networks}
%%%%%%%%%%%%%%%%%%%%%%%%%%%%%%%%%%%%%%%%%%%%%%%%%%%%%%%%%%%%%%%%%
A fundamental property of variational quantum circuits with data-encoding rotations is their intrinsic Fourier structure~\cite{reupload,schuld}. When classical variables are encoded into parameterised quantum rotations, the expectation values of quantum observables can be expressed as finite Fourier series. The accessible frequencies are determined by the encoding operations, whereas the corresponding Fourier coefficients depend on the trainable circuit parameters. Repeated encoding of the input variables, commonly referred to as data re-uploading, extends the accessible frequency spectrum and enhances the expressivity of the quantum model. This structure provides a useful framework for numerical integration. Instead of directly integrating a complicated function, the QNN is trained to approximate the target integrand through a finite combination of trigonometric functions. The resulting Fourier representation can then be integrated term by term using quantum amplitude estimation.

For the two-dimensional NLO decay-rate integrands obtained in Sec.~\ref{sec:ltd}, we employ the six-qubit QNN architecture introduced in Ref.~\cite{original}. The two integration variables are encoded through $R_x$ rotations, each repeated across three qubits. The encoding operations are interleaved with trainable rotations and entangling gates, while the quantum circuit output is obtained from measurements in the Pauli-$Z$ basis. The QNN is implemented using PennyLane and trained with the Adam optimiser. The numerical setup employs 20 layers, a learning rate of $10^{-3}$ and up to 15,000 optimisation steps. Following Ref.~\cite{original}, the Fourier representation is truncated to 20 retained coefficients. The accuracy of the Fourier approximation depends on the ability of the QNN to reproduce the original decay-rate integrand. Therefore, the regression error introduced during the Fourier reconstruction must be distinguished from the uncertainty associated with the subsequent quantum integration.

%%%%%%%%%%%%%%%%%%%%%%%%%%%%%%%%%%%%%%%%%%%%%%%%%%%%%%%%%%%%%%%%%
\subsection{Quantum integration and numerical results}
%%%%%%%%%%%%%%%%%%%%%%%%%%%%%%%%%%%%%%%%%%%%%%%%%%%%%%%%%%%%%%%%%

\begin{figure}[t]
\centering
\includegraphics[width=0.9\textwidth]{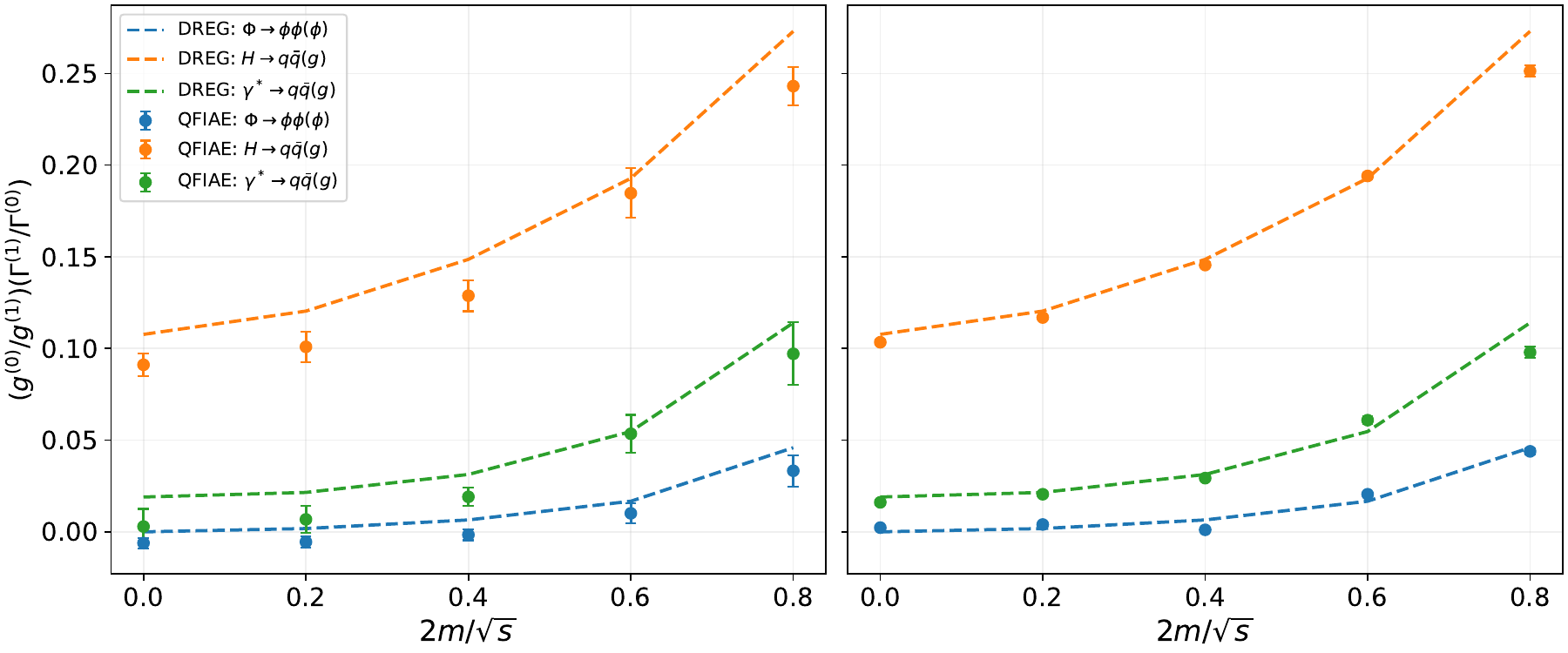}
\caption{
NLO decay rates for $H\to q\bar q(g)$,
$\gamma^*\to q\bar q(g)$ and
$\Phi\to\phi\phi(\phi)$ as functions of the
final-state mass ratio $r=2m/\sqrt{s}$.
Left: QFIAE results obtained with both the QNN
and IQAE modules executed on quantum simulators.
Right: hybrid implementation with the QNN
executed on a simulator and IQAE performed
on quantum hardware. Dashed curves represent
the analytical predictions obtained in
dimensional regularisation (DREG).
The QNN employs six qubits and 20 layers,
while the IQAE module uses five
qubits~\cite{original}.}
\label{fig:results}
\end{figure}

Once the QNN has been trained, the Fourier coefficients are extracted from the quantum circuit and the corresponding trigonometric contributions are integrated using Iterative Quantum Amplitude Estimation (IQAE)~\cite{iqae,original}. IQAE estimates the relevant quantum amplitudes through repeated measurements and classical post-processing, without requiring quantum phase estimation. The individual integrals are combined with their Fourier coefficients to reconstruct the total NLO decay rate. The IQAE implementation uses five qubits, $10^3$ measurement shots, a target accuracy $\epsilon=0.01$ and a confidence parameter $\alpha=0.05$ for each individual integral. Further details of the quantum integration procedure are given in Refs.~\cite{iqae,original}. Figure~\ref{fig:results} presents the NLO decay rates for $H\to q\bar q(g)$, $\gamma^*\to q\bar q(g)$ and $\Phi\to\phi\phi(\phi)$ as functions of the final-state mass ratio $r=2m/\sqrt{s}$. The numerical results are compared with the corresponding analytical predictions obtained in dimensional regularisation (DREG).

The quantum simulator results reproduce the overall mass dependence of the three decay channels and show reasonable agreement with the analytical predictions. Nevertheless, some deviations exceed the reported quantum-integration uncertainties, indicating that the accuracy of the Fourier approximation and that of the integration algorithm should be assessed separately. The hardware implementation exhibits larger uncertainties and generally lower central values than the corresponding DREG predictions. These deviations reflect the limitations of noisy quantum devices and are not fully removed by the error-mitigation procedures employed in Ref.~\cite{original}. Importantly, the hardware execution concerns the IQAE module, whereas the QNN is trained and evaluated on a quantum--simulator. The results demonstrate the feasibility of combining a QNN-based Fourier representation with quantum amplitude estimation to evaluate physical decay rates at NLO. Although a computational quantum advantage is not established, this approach provides a proof of concept for the application of quantum neural networks to perturbative high-energy physics calculations.

%%%%%%%%%%%%%%%%%%%%%%%%%%%%%%%%%%%%%%%%%%%%%%%%%%%%%%%%%%%%%%%%%
%%%%%%%%%%%%%%%%%%%%%%%%%%%%%%%%%%%%%%%%%%%%%%%%%%%%%%%%%%%%%%%%%
\section{Conclusions}

We have explored quantum neural networks (QNNs) as a tool for the numerical integration of physical observables in high-energy physics. By combining the locally finite integrands provided by causal-unitary LTD with the Quantum Fourier Iterative Amplitude Estimation (QFIAE) algorithm, we have evaluated decay rates at next-to-leading order (NLO). The central feature of this approach is the intrinsic Fourier structure of parameterised quantum circuits, which allows the QNN to approximate the integrands through truncated Fourier series whose coefficients are determined by training. The resulting trigonometric contributions are subsequently integrated using Iterative Quantum Amplitude Estimation (IQAE).

The results for three benchmark decay processes reproduce the overall mass dependence of the analytical NLO predictions, with better agreement in quantum simulations than in the hardware implementation. The accuracy of the method depends on the quality of the Fourier approximation, the number of retained coefficients and the uncertainties introduced during quantum integration. Together with hardware noise and computational resource requirements, these limitations must be investigated before establishing any practical quantum advantage.

More broadly, the ability of QNNs to learn and represent nontrivial functions motivates their application beyond perturbative calculations. In particular, quantum circuit models have recently been investigated for the representation of hadron fragmentation functions~\cite{renteriaFF}, opening opportunities to explore quantum machine learning in the long-distance, non-perturbative sector of QCD. These developments highlight the potential of QNNs as flexible computational tools for both perturbative and non-perturbative problems in high-energy physics.

\paragraph{Acknowledgments.} \sloppy This work is supported by the Spanish Government and ERDF/EU - Agencia Estatal de Investigaci\'on (MCIN/AEI/10.13039/\allowbreak 501100011033) Grants No.\ PID2023-146220NB-I00 and CEX2023-001292-S, by Generalitat Valenciana Grants No.\ CISPINCOM/2026/004 and CIGRIS/2022/145, by the Quantum Spain project and by the CSIC Interdisciplinary Thematic Platform on Quantum Technologies (PTI-QTEP+).


\begin{thebibliography}{20}\small
\bibitem{schuld} M.~Schuld, R.~Sweke and J.~J.~Meyer, \emph{Effect of data encoding on the expressive power of variational quantum-machine-learning models}, Phys. Rev. A \textbf{103} (2021) 032430.

\bibitem{reupload} A.~P\'erez-Salinas et al., \emph{Data re-uploading for a universal quantum classifier}, Quantum \textbf{4} (2020) 226.

\bibitem{qfiae} J.~J.~Mart\'inez de Lejarza, M.~Grossi, L.~Cieri and G.~Rodrigo, \emph{Quantum Fourier iterative amplitude estimation}, 2023 IEEE International Conference on Quantum Computing and Engineering, pp.~571--579; arXiv:2305.01686.

\bibitem{Pyretzidis:2025stx} K. Pyretzidis, J.~J.~Mart{\'\i}nez de Lejarza and G. Rodrigo, \emph{Unlocking multidimensional integration with quantum adaptive importance sampling}, Commun. Phys. 9 (2026) 1, 269, arXiv:2506.19965.

\bibitem{original} J.~J.~Mart\'inez de Lejarza, D.~F.~Renter\'ia-Estrada, M.~Grossi and G.~Rodrigo, \emph{Quantum integration of decay rates at second order in perturbation theory}, Quantum Sci. Technol. 10 (2025) 2, 025026, arXiv:2409.12236.

\bibitem{vacuum} S.~Ram\'irez-Uribe et al., \emph{Rewording theoretical predictions at colliders with vacuum amplitudes}, Phys. Rev. Lett. 133 (2024) 21, 21, arXiv:2404.05491.

\bibitem{causal} LTD Collaboration, \emph{Vacuum amplitudes and time-like causal unitary in the loop-tree duality}, JHEP 01 (2025) 103, arXiv:2404.05492.

\bibitem{iqae} D.~Grinko, J.~Gacon, C.~Zoufal and S.~Woerner, \emph{Iterative quantum amplitude estimation}, npj Quantum Inf. \textbf{7} (2021) 52.

\bibitem{renteriaFF} D.~F.~Renter\'ia-Estrada, R.~J.~Hern\'andez-Pinto, G.~Rodrigo, R.~Sassot and G.~F.~R.~Sborlini, \emph{A quantum representation of $\pi$ fragmentation functions through variational quantum circuits}, arXiv:2609.20557.
\end{thebibliography}
\end{document}